\documentclass{article}
\usepackage[ansinew]{inputenc}
\usepackage{latexsym}
\newcommand{\be}{\begin{equation}}
\newcommand{\ee}{\end{equation}}
\newcommand{\bear}{\begin{eqnarray}}
\newcommand{\ear}{\end{eqnarray}}
\newcommand{\ba}{\begin{eqnarray*}}
\newcommand{\ea}{\end{eqnarray*}}
\renewcommand{\theequation}{\arabic{section}.\arabic{equation}}

\newcommand{\no}{\noindent}

\newcommand{\os}{\mbox{\tiny{1-S}}}

\newcommand{\limj}{\lim_{f\rightarrow 1,L\rightarrow\infty}}

\begin{document}

\date{\today}
\title{\textbf{Charge Fluctuations in Soliton Anti-Soliton Systems Without Conjugation Symmetry}}
\author{R. L. P. G. Amaral$^{a}$, V. E. R. Lemes$^{b}$ and O.S. Ventura$^{b,c}$
\footnote{rubens@if.uff.br, vitor@dft.if.uerj.br and ozemar@ifes.edu.br}\\
{\small\em $^a$Instituto de F\'\i sica, Universidade Federal Fluminense, 24210 - 340, Niter\'oi - RJ, Brasil,}\\
\small\em $^b$Instituto de F\'\i sica, Universidade do Estado do Rio de Janeiro,\\
\small\em Rua S\~{a}o Francisco Xavier 524, Maracan\~{a}, Rio de Janeiro - RJ, 20550-013, Brazil \\
\small \em $^c$Coordenadoria de F\'{\i}sica, Instituto Federal do Esp\'{\i}rito Santo,  \\
\small\em $ $Avenida Vit\'{o}ria 1729 - Jucutuquara, Vit\'{o}ria - ES, 29040 - 333, Brazil\\ }
\bigskip
\maketitle

\maketitle
\begin{abstract}
We analyse de fluctuations of charge of the (1+1)D Dirac´s fermion with charge conjugation breaking. This is done taking the separation between background soliton and anti-soliton going to infinity.
\end{abstract}

\section{Introduction}

The interest in the investigation of non-integer charge has become stronger in quantum field theory since the original work \cite{gj}. This happened because this charge conjugation symmetric (1+1)D model, with a soliton anti-soliton background, predicts $\frac e2$ as an eigenvalue of the charge operator and has been used to describe properties of the polyacetilen\cite{S,NS}. The concept of fractionary charge has been reinforced when it was shown that there is no charge fluctuations
\cite{JKKS,RB}. These ideas opened a door to investigations in condensed matter using, as a tool, quantum field theory.

The possibility that some system could have non-integer charge was also investigated in a variety of other contexts. Excitations of charge condensates were studied by
Rice at all \cite{RICE}. More recently, there has been renewed interest for non-integer quantum numbers. The construction of high quality samples of grapheme made it possible the discussion of two-spatial dimensional field theory models applied to these materials. Distortions of the honeycomb lattice of grapheme lead, in the continuum, to theories with fermions coupled to scalar fields (complex Higgs fields) and their topological excitations such as vortices induces mid-gap states. This state of affair reproduces much of the excitement previously experienced with polyacetilene in this higher dimensional analog. In particular, charge fractionalization through the breaking of energy-reflection symmetry reappears \cite{JPI,HOCH,CHAHOJA}. This mechanism should be contrasted with the previously studied phenomenon of fractional charges as in the fractional Hall effect, in two spatial dimensions , which rely on the forced breaking of time-reversal symmetry associated to topological order. This new mechanism is much closer to the ones witnessed with polyacetilene. Instead of fractional charges generic values for the charge are obtained.

It is very natural that variations of the original models predict a much richer spectrum of possibilities. The simplest generalization is to include breaking of charge conjugation in the Jackiw’s first proposal \cite{JKKS} as done in \cite{Jackiw2} who obtained general non-integer charges. Further investigations have been done \cite{CAPRI1, CAPRI2}. In \cite{gw} a general method for computing the charge has been applied to two- and four- dimensional models.

However it is missing, up to now, the investigation of the fluctuations of the charge within models without charge conjugations symmetry. To perform this investigation within  the simplest case is the goal of this work.

In section 2, we present the model and construct the localized charge operator. In following section we compute the charge eigenvalue. The fluctuations are computed in section 4.

\section{The number operator}

Let us start  with the (1+1) dimensional Dirac equation
$$\label{1}
E\Psi=(\sigma_2 p+\sigma_1\phi+\gamma \sigma_3)\Psi.
$$
 In terms of components,  the equations of motion are

\begin{eqnarray}\label{components}
(-ip+\phi)\psi_2&=&(E-\gamma)\psi_1,\nonumber\\
(ip+\phi)\psi_1&=&(E+\gamma)\psi_2.
\end{eqnarray}
Here $p=-id/dx$, $\phi(x)$ presents an soliton-anti-soliton background profile

\begin{displaymath}
\label{defsol} \phi(x)= \left\lbrace
\begin{array}{ll}
   +\phi_{0} &, x<0 \\
   -\phi_{0} &, 0\leq x \leq L \\
   +\phi_{0} &, L<x\\
  \end{array} \right.
\end{displaymath}
and $\gamma$ is a constant parameter responsible for the breaking of the Hamiltonian conjugation symmetry.

\noindent The continuum energies are given by

$$E=\pm \sqrt{k^2+\gamma^2+\phi_0^2},$$
and the positive energy solutions can be written as

$$\label{solution1}
\Psi^+_k(x)=\left(\begin{array}{c}
\theta^+u_k(x)\\ \theta^-v_k(x),
\end{array}\right)
$$
while the negative ones as

$$\label{solution2}
\Psi^-(x)_k=\left(\begin{array}{c}
\theta^-u_k^*(x)\\ -\theta^+v_k^*(x)
\end{array}\right),
$$
\no where

$$\theta^\pm=\theta^\pm(k)=\sqrt{\frac{|E|\pm\gamma}{|E|}}.$$

\noindent It is remarkable that there is no zero-energy continuum, i.e., $\left|E\right|\geq\sqrt{\gamma^2+\phi_0^2}$. For the bound states, the analogous occur, i.e., the positive energy solution is given by

$$\label{solutionb1}
\Psi_B^+(x)=\left(\begin{array}{c}
\theta^+_Bu_B(x)\\ \theta^-_Bv_B(x)
\end{array}\right),
$$
and the negative one is written as
$$\label{solutionb2}
\Psi_B^-(x)=\left(\begin{array}{c}
\theta^-_Bu_B(x)\\ -\theta^+_Bv_B(x)
\end{array}\right),
$$
where the parameters $\theta^\pm_B$ are defined in terms of the the bound-state energy in  analogy to definitions of $\theta^\pm(k)$. Note that thes parameters are equal to 1 when $\gamma$ is iqual to zero, so that the functions $u$´s and $v$´s become the upper and lower compontents of $\Psi$. In this presentation the functions $u_k(x)$, $v_k(x)$, $u_B(x)$ and $v_B(x)$ satisfy the eq. (\ref{components}) with $\gamma=0$. All contribution of the conjugation breaking is codified by the parameters $\theta^\pm$. The bound states energies are $ E_B=\pm\sqrt{-\kappa^2+\phi_0^2+\gamma^2}$, where the parameter $\kappa$ is defined by a transcendental equation which does not depend on $\gamma$,

$$
\phi_0^2-\kappa^2=\phi_0^2\exp{(-2\kappa L)}.
$$
The normalizations are

\begin{eqnarray}\label{norm1}
\int_xu_k^*u_k=\int_xv_k^*v_k&=&\frac 12\delta(k-k^\prime),\\
\label{norm2}\int_xu_B^*u_B=\int_xv_B^*v_B&=&\frac 12
\end{eqnarray},
and the overlaps between $u_B$ and $u_k$ as well as between $v_B$ and $v_k$ vanish.

Observe that the "conjugation" transformation

\be
\Psi\longrightarrow (c_1\sigma_3+c_2)\Psi^\dagger
,\ee

\no with
\be
c_1={1\over 2}\left(\frac{{\theta^+}^2+{\theta^-}^2}{\theta^+\theta^-}\right)
\ee
\noindent and
\be
c_2={1\over 2}\left(\frac{{\theta^+}^2-{\theta^-}^2}{\theta^+\theta^-}\right),
\ee
maps positive to negative energy states.
It is important to note that, in spite of the conjugation mapping above, $\rho_k^{E>0}\neq\rho_k^{E<0}$.
In the present case, defining the first quantized scalar and chiral densities $\rho=\Psi^\dagger\Psi$ and $\widetilde\rho=\Psi^\dagger\sigma^3\Psi$,

$$
\rho_{-k}^{E<0}=({c_1}^2+{c_2}^2)\rho_k^{E>0}+2c_1c_2\widetilde\rho_k^{E>0}.
$$
This implies that the fermion charges acquires a generic value, not restricted to half-integer values \cite{Jackiw-primeiraquantizacao}.

The second quantized quantum field operator can be expanded as

\be \label{operator}
\Psi =\int \!\!\!\!\!\!\!\sum_k \left(e^{-iE_kt}\left(\begin{array}{c}
\theta^+u_k(x)\\ \theta^-v_k(x)
\end{array}\right)a_k+e^{iE_kt}\left(\begin{array}{c}
\theta^-u_k^*(x)\\ -\theta^+v_k^*(x)
\end{array}\right)b_k^\dagger\right)
\ee
with canonical anticommutation relations between $a_k$, $b_k$, $a_B$ and $b_B$, i.e.,

\bear
\left\lbrace a_{k}, a_{k'}^{\dagger}\right\rbrace & =& \left\lbrace b_{k}, b_{k'}^{\dagger}\right\rbrace = \delta (k-k'), \nonumber \\
\left\lbrace a_{B}, a_{B}^{\dagger}\right\rbrace & =& \left\lbrace b_{B}, b_{B}^{\dagger}\right\rbrace = 1.
\label{com-rel}
\ear
If we define the localized number operator as

\be
N_f=\int_x \frac f2\left[\Psi^\dagger,\Psi\right]
,\ee
where $f(x)$ plays the role of an smearing function, peaked near the soliton, we obtain

\bear
N_f&=&\sum\!\!\!\!\!\!\!\!\!\int_{x,k,k^\prime} f\left[e^{-i(E_k-E_{k^\prime})t}\left[a_k^\dagger a_k(\theta^+\theta^{\prime +}
u_{k^\prime}^*u_k+\theta^-\theta^{\prime -}
v_{k^\prime}^*v_k)-b_k^\dagger b_k(\theta^-\theta^{\prime -}
u_{k^\prime}^*u_k+\theta^+\theta^{\prime +}
v_{k^\prime}^*v_k)\right]\right.\nonumber\\
&&\hskip 2cm\left.+e^{-i(E_k+E_{k^\prime})t}\left[b_k
a_k(\theta^+\theta^{\prime -} u_{k^\prime}u_k-\theta^-\theta^{\prime
+} v_{k^\prime}v_k)\right]\right.\nonumber\\&& \hskip
2cm\left.+e^{+i(E_k+E_{k^\prime})t}\left[a_k^\dagger
b_k^\dagger(\theta^-\theta^{\prime +}
{u_{k^\prime}}^*u_k^*-\theta^+\theta^{\prime
-}{v_{k^\prime}}^*v_k^*)\right]\right.\nonumber\\&& \hskip
2cm\left.+\frac
12\delta(k-k^\prime)\left[({\theta^-}^2-{\theta^+}^2)u_k^*u_k-({\theta^-}^2-{\theta^+}^2)v_k^*v_k\right]
\right], \ear
which can be expressed as

\be
N_f=\hat N_B^f+ N_C^f+\hat N_k^f. \label{eq12a} \ee
Here $N_B^f$ represents the quantum contributions of the bound state sector, $N_k^f$ the
continuum sector and $N_C$ is the c-number contribution coming from both
the bound as well as the continuum sector after subtractions of the zero soliton background.

\bear
\hat N_B^f&=&\int_{x} f\left[\left(a_B^\dagger a_B({\theta_B^+}^2
u_{B}^2+{\theta^-_B}^2 v_{B}^2)-b_B^\dagger b_B({\theta^-_B}^2
u_{B}^2+{\theta^+_B}^2
v_{B}^2)\right)\right.\nonumber\\
&&\left.+e^{-2iE_Bt}\left(b_k a_k({\theta^+_B}\theta^-_B
u_{B}^2-{\theta^-_B}\theta^+_B v_{B}^2)\right)\right.\nonumber\\&&
\left.+e^{+i2E_Bt}\left(a_B^\dagger b_B^\dagger(\theta^-_B\theta^{
+}_B {u_{B}^2}-\theta^+_B\theta^{-}_B{v_{B}^2})\right)\right],
\label{eq113}\ear

\bear
N_C^f&=&\int_{x,k} \frac f2\left[({\theta^-}^2-{\theta^+}^2)(u_k^*u_k-v_k^*v_k)-({\theta^-_0}^2-{\theta^+_0}^2)({u_k^0}^*{u_k^0}-{v_k^0}^*{v_k^0})\right]\\
&&+\int_x \frac f2\left[({\theta^-_B}^2-{\theta^+_B}^2)(u_B^2-v_B^2)\right],
\ear
where the subscript $0$, in $\theta^\pm_0$, means that the energies are the ones with no soliton background, but with $\gamma\neq 0$ and
$$
\hat N_k^f=\hat N_k^{f+-}+\hat N_k^{f++}+\hat N_k^{f--}.
$$
Here it is useful to decompose the continuum contribution in terms of the creation and annihilation operators

 \be \hat N_k^{f+-}=\int_{x,k,k^\prime}
f\left[e^{-i(E_k-E_{k^\prime})t}\left(a_{k^\prime}^\dagger
a_k(\theta^+\theta^{\prime +}
u_{k^\prime}^*u_k+\theta^-\theta^{\prime -}
v_{k^\prime}^*v_k)-b_{k^\prime}^\dagger b_k(\theta^-\theta^{\prime
-} u_{k^\prime}^*u_k+\theta^+\theta^{\prime +}
v_{k^\prime}^*v_k)\right)\right], \ee
\be \hat
N_k^{f--}=\int_{x,k,k^\prime}
f\left[e^{-i(E_k+E_{k^\prime})t}\left(b_{k^\prime}
a_k(\theta^+\theta^{\prime -} u_{k^\prime}u_k-\theta^-\theta^{\prime
+} v_{k^\prime}v_k)\right)\right] \ee
 and
 \be
\hat N_k^{f++}=\int_{x,k,k^\prime}
f\left[e^{+i(E_k+E_{k^\prime})t}\left(a_{k^\prime}^\dagger
b_k^\dagger(\theta^-\theta^{\prime +}
{u_{k^\prime}}^*u_k^*-\theta^+\theta^{\prime
-}{v_{k^\prime}}^*v_k^*)\right)\right]. \ee
In the last  equations,  $\theta^\prime=\theta(k^\prime)$.

It is important to note that there is a contribution to the c-number term coming from the continuum states which appears when  $\gamma\neq0$.
It is not eliminated by subtracting the vacuum contributions computed in the zero soliton sector. Indeed, although we have explicited these zero soliton regularizing terms they turn out to be identically zero, since $|u_k^0|^2=|v_k^0|^2$. These regularizing terms are necessary if we compute  separately the
contributions from $u_k$ and  $v_k$.

It is convenient to rewrite these equations by defining the symmetrized and anti-symmetrized combinations of the factors defined below

$$
\theta^{(1)}_S=(\theta^-(k)\theta^+(k^\prime)+\theta^+(k)\theta^-(k^\prime))/2,
$$
$$
\theta^{(1)}_A=(\theta^-(k)\theta^+(k^\prime)-\theta^+(k)\theta^-(k^\prime))/2,
$$
and
$$
\theta^{(2)}_S=(\theta^+(k)\theta^+(k^\prime)+\theta^-(k)\theta^-(k^\prime))/2,
$$
$$
\theta^{(2)}_A=(\theta^+(k)\theta^+(k^\prime)-\theta^-(k)\theta^-(k^\prime))/2.
$$
With these definitions we obtain

\be \hat N_k^{f+-}=\int_{x,k,k^\prime}
f\left[e^{-i(E_k-E_{k^\prime})t}\left[(a_{k^\prime}^\dagger
a_k-b_{k^\prime}^\dagger b_k)\theta^{(2)}_S(
u_{k^\prime}^*u_k+v_{k^\prime}^*v_k)+(a_{k^\prime}^\dagger
a_k+b_{k^\prime}^\dagger b_k)\theta^{(2)}_A(
u_{k^\prime}^*u_k-v_{k^\prime}^*v_k)\right]\right], \ee \be \hat
N_k^{f--}=\int_{x,k,k^\prime}
f\left[e^{-i(E_k+E_{k^\prime})t}b_{k^\prime}
a_k\left[\theta^{(1)}_S(
u_{k^\prime}u_k-v_{k^\prime}v_k)-\theta^{(1)}_A(u_{k^\prime}u_k+
v_{k^\prime}v_k)\right]\right] \ee and \be \hat
N_k^{f++}=\int_{x,k,k^\prime}
f\left[e^{+i(E_k+E_{k^\prime})t}b_{k^\prime}^\dagger
a_k^\dagger\left[\theta^{(1)}_S(
u_{k^\prime}^*u_k^*-v_{k^\prime}^*v_k^*)-\theta^{(1)}_A(u_{k^\prime}^*u_k^*+
v_{k^\prime}^*v_k^*)\right]\right]. \ee
On the next section we will investigate the fluctuations of the eigenvalues of this operator.

\section{Charge fluctuations}

Let us first remark here that as $f\longrightarrow 1$, $N_{f}$ becomes diagonal, with integer eigenvalues. This corresponds to the number operator of the model with the
soliton  anti-soliton distance fixed. In this case the orthonormality
of the functions can be invoked, leading to

$$
\hat N_k^f\longrightarrow \hat N_k=\int_k (a_k^\dagger a_k-b_k^\dagger b_k)
,$$

$$
N_B^f\longrightarrow N_B=a^\dagger_Ba_B-b^\dagger_Bb_B
$$
and

$$
N_C^f\longrightarrow 0.
$$

Now let us turn to the study of the limit $L\longrightarrow \infty$ followed by $f\longrightarrow 1$.
Since the $u_k(x)$ and $v_k(x)$ functions are  independent of $\gamma$, we have

\be \lim_{f\rightarrow
1,L\rightarrow\infty}\int_xf(u^2_B+v^2_B)=\frac 12 \label{eq1}\ee
and \be \lim_{f\rightarrow
1,L\rightarrow\infty}\int_xf(u^2_B-v^2_B)=-\frac 12 ,\label{eq2}\ee
where $L$ is the soliton-antisoliton distance.

In the following we use the point-wise limit

\be \lim_{ L\rightarrow \infty} |u_k|^{2}= |u_k^{\os}|^{2}
\label{eq3},\ee
 and
 \be \lim_{ L\rightarrow \infty} |v_k|^{2}=
|v_k^{\os}|^{2} \label{eq4},\ee
with
 \be |\lim_{
L\rightarrow \infty}  u_k|^{2}\neq  \lim_{ L\rightarrow \infty}
|u_k|^{2} \label{eq5},\ee
 where $u_k^{\os}$ and
$v_k^{\os}$ corresponds to the solutions, with $\gamma=0$ and
with only one anti-soliton and no soliton. See the appendix A for
the proof of equations
(\ref{eq3} -  \ref{eq5}).

The limit of the bound-state energies, in the long distance case, is characterized  by

$$\lim_{L\rightarrow \infty} E_B= \gamma.$$
In this limit we obtain that

$$
\lim_{f\rightarrow 1,L\rightarrow\infty}
\left(\begin{array}{c}
u_B \\
v_B
\end{array}\right) =\left(\begin{array}{c}
0\\
v_B^{\os}
\end{array}\right),
$$
and that

$$\lim_{f\rightarrow 1,L\rightarrow\infty} \theta_B^+=\sqrt{2},\   \  \lim_{f\rightarrow 1,L\rightarrow\infty} \theta_B^-=0.$$
From this we conclude that  there is one state with negative energy $E=-\gamma$ in the limiting case.

The situation where $L\rightarrow\infty$ followed by $f\rightarrow 1$ is caracterized by some further usefull properties, i.e.,

\be \lim_{f\rightarrow 0,L\rightarrow\infty} \int_{x}f  \left(
u_{k^\prime}^*u_{k}-v_{k^\prime}^*v_{k}\right)= 0, \label{ea1}\ee

\be \limj \int_{x}f  \left(
u_{k^\prime}u_{k}-v_{k^\prime}v_{k}\right)= 0, \label{ea2}\ee

\be \limj \int_{x}f  \left(
u_{k^\prime}^*u_{k}+v_{k^\prime}^*v_{k}\right)= T\delta
(k-k^{\prime})+ R\delta (k+k^{\prime}) \label{ea3}\ee
and
\be \limj \int_{x}f  \left(
u_{k^\prime}u_{k}+v_{k^\prime}v_{k}\right)= T^{\prime}\delta
(k+k^{\prime})+ R^{\prime}\delta (k-k^{\prime}). \label{ea4}\ee
Using these relations and the explicit form of $\theta^{\pm}$ we
obtain

$$
\hat{N}^{f + -}_{k}\longrightarrow \int_{k}\left(T(k)(a_{k}^{\dagger}a_{k} -
b_{k}^{\dagger}b_{k}) + R(k)(a_{-k}^{\dagger}a_{k} -
b_{-k}^{\dagger}b_{k})\right)
,$$

$$ \hat N _{k}^{f,\pm\pm}\longrightarrow 0   $$
and
$$ \hat N_B^f\longrightarrow N_B^\infty=-b^\dagger_Bb_B,
$$
so that

\be N_C^f\longrightarrow\int_{x,k} \frac
12\left[({\theta^-}^2-{\theta^+}^2)
{u_k^{\os}}^*u_k^{\os}-({\theta^-}^2-{\theta^+}^2){v_k^{\os}}^*v_k^{\os}
\right] +\frac 12.\ee
As has been shown in the appendix B, we obtain that

$$
N_C^f\longrightarrow \frac{-1}{\pi}\tan^{-1}{({\phi_0\over\gamma})}
.$$
Note that the limit when $\gamma$ goes to zero leads to the
$-\frac{1}{2}$ value expected. The same result is obtained by
considering  the case when we start with only one anti-soliton background, see appendix B.

It is interestingly to note that the investigation of the case $\gamma\neq0$  easier to
deal than the case $\gamma =0$. This occurs since there is no mixing between the creation-destruction
 operators for the $N_B$ term
in the limit $L$ going to infinite and $f$ going to one. In contrast to $\gamma = 0$ case, no  Bogoliubov transformation is necessary.
Let us compute
the fluctuations to the state

$$
| \psi > = b^\dagger |0,\bar 0> = |0,\bar 1>.
$$
Using eq. (\ref{eql2a}), we obtain

\bear\limj <\psi|{N}^{f}|\psi> &=& \limj
<\psi|\hat{N}_{k}^{f}|\psi> + \limj <\psi|\hat{N}_{B}^{f}|\psi> +
\limj <\psi|{N}_{C}^{f}|\psi> \nonumber \\
&=& 0 + 1 + N_{C}^{\infty}. \label{autval} \ear
Let´s consider the quadratic number operator

\be
(N^{f})^{2}=(\hat N_{k}^{f})^{2}+(\hat N_{B}^{f})^{2}+(N_{C}^{f})^{2} + 2
\hat{N}_{k}^f\hat{N}_{B}^f + 2{N}_{C}^{f}(\hat{N}_{k}^{f}+
\hat{N}_{B}^{f}).
\ee
So we have to compute

\bear \limj <\psi|(N_{k}^{f})^{2}|\psi> &=& \limj <\psi|(
\hat{N}_{k}^{f,+-} + \hat{N}_{k}^{f,++} +
\hat{N}_{k}^{f,--})^{2})|\psi> \nonumber \\
&=& \limj <\psi|( \hat{N}_{k}^{f,--} \cdot \hat{N}_{k}^{f,++})|\psi>
\nonumber \\
&=& \limj  \int_{k,k^{'},x,y}f_{x}f_{y}\{ (\theta_{S}^{(1)})^{2}
(u_{k}u_{k^{'}} - v_{k}v_{k^{'}})_{x}(u_{k}^{\ast}u_{k^{'}}^{\ast} -
v_{k}^{\ast}v_{k^{'}}^{\ast})_{y} \nonumber \\
&&
\hspace{3cm}-\theta_{A}^{(1)}\theta_{S}^{(1)}(u_{k}^{\ast}u_{k^{'}}^{\ast}
+ v_{k}^{\ast}v_{k^{'}}^{\ast})_{y}(u_{k}u_{k^{'}} +
v_{k}v_{k^{'}})_{x} \nonumber \\
&&
\hspace{3cm}-\theta_{A}^{(1)}\theta_{S}^{(1)}(u_{k}^{\ast}u_{k^{'}}^{\ast}
+ v_{k}^{\ast}v_{k^{'}}^{\ast})_{y}(u_{k}u_{k^{'}} +
v_{k}v_{k^{'}})_{x} \nonumber \\
&&\hspace{3cm}+ (\theta_{k}^{(1)})^{2}(u_{k}^{\ast}u_{k^{'}}^{\ast}
+ v_{k}^{\ast}v_{k^{'}}^{\ast})(u_{k}u_{k^{'}} + v_{k}v_{k^{'}}) \}.
\label{eqqua} \ear
The first term in the last equation is discussed in \cite{JKKS} and is
seen to be 0 in this limit. For the two middle terms the integrals
of the factors with negative signs between $u^{\ast}u$ and
$v^{\ast}v$ vanish for analogous reasons. That the last term also
vanishes comes from (\ref{ea1}-\ref{ea4}) which
makes $\theta_{A}^{(1)}\longrightarrow 0$. The result is
$$
\limj <\psi|(\hat{N}_{f}^{+})^{2}|\psi>=0.
$$
Furthermore
$$
\limj <\psi|(\hat{N}_{B}^{+})^{2}|\psi>=1
$$
and
$$
\limj (N_{C}^{f})^{2} = (N_{C}^\infty )^{2}.
$$
Also
$$
\limj <\psi|(\hat{N}_{k}^f\hat{N}_{B}^f)|\psi>=0
$$
and
$$ \limj <\psi|2N_{C}^{f}(\hat{N}_{k}^f + \hat{N}_{B}^f)|\psi> =
2N_{C}^{\infty}.
$$
Collecting all terms we obtain $$ <\psi|(\hat{N}^{+})^{2}|\psi> =
(N_{C}^{\infty})^{2} + 2 N_{C}^{\infty} + 1 = (N_{C}^{\infty} +
1)^{2},$$
so that
$$<\psi|\delta(\hat{N}^{+})^{2}|\psi>=0,$$
which means that there is no fluctuations.

\section{Conclusions and Discussion}

In this paper, we have studied the charge eigenvalues and their fluctuations for a model without charge symmetry that describes fermions, in (1+1)D in a soliton anti-soliton background.  This has been done in the context of second quantization extending the analysis of \cite{Jackiw2} to the case of soliton and anti-soliton. Although this model had been throughously  investigated in the literature it was remaining to be shown that the variance of charge expected value  vanishes, which is a necessary condition to imply that this charge is a good quantum number. We have obtained zero fluctuations as the result of our analysis.

It is interesting to note that in the case $\gamma\neq 0$ there appears a C-number contribution to the charge operator that misses in the case $\gamma=0$.
This contribution, which comes from the continuum states,
  turns out to be  responsible for the existence of the continuum spectrum of the charge when $\gamma\neq  0$, varying between $\pm { 1\over 2}$.  If we take a look in the resultant charge, eq. (\ref{v44}), we  identify a part that is ${\frac 1 2}$ and another that is continuous. The second part collects the contributions from the charge conjugation breaking term. Curiously, in a certain sense, the treatment is  easier when $\gamma \neq 0$. This occurs since in the presence of a soliton and an anti-soliton there are two finite-norm fermionic states centered in the soliton or the anti-soliton positions. Their energies go to zero when the distance grows for the case
 $\gamma=0$ but remain distinct in the case $\gamma\neq 0$. This makes the overlapping of the solutions in the infinite distance limit finite when $\gamma=0$ and leads to the necessity of a Bogoliubov transformation that is  not necessary when $\gamma\neq 0$.

 The strategy of this work has been to inquire about the physical meaning of the charge induced by the soliton through the imersion of the physical situation in a broader one where a soliton and anti-soliton are present. Then the simplest case should be re-obtained in the limit when the soliton to anti-soliton distance grows infinitely. If the
 fluctuations of the charge operator defined around the soliton were kept finite this would be a strong evidence of the partial character of the non-integer charge. In this
 analysis the point-wise limits eqs. (\ref{eq3} -  \ref{eq5}) have played an important role. They contain the information that those systems, soliton anti-soliton and only a soliton background, are equivalent after we take the limit when the distance between soliton and anti-soliton goes to infinity.
 We have obtained these relations in the sense of distribution theory and have taken strict control on the orthonormality character of the functions involved in their definitions along the limiting process.

  Lets us stress that the lessons of our analysis should be applicable to field theory models of graphene which present fermion fractionalization  in 2+1D also
  without conjugation symmetry.

\section*{Acknowledgments}

We  would like to thank R. Jackiw for suggesting the problem and for  fruitful discussions as well as the Center for Teoretical Physics / MIT and Departamento de Física Teórica da UERJ for the hospitality given to  O.S.Ventura  when this work has been done.  O.S.Ventura was supported by Funda\c c\~ao  de Amparo \`a Pesquisa do Estado do Rio de
Janeiro (Faperj), Conselho Nacional de Desenvolvimento
Cient\'ifico e Tecnol\'ogico (CNPq-Brazil) and also, in the beginning of this work, by
FUNCEFETES and Pr\'o-Reitoria de Pesquisa e P\'os-gradua\c c\~ao at Ifes.
The work of V.E.R.Lemes was supported by Conselho Nacional de Desenvolvimento
Cient\'ifico e Tecnol\'ogico (CNPq-Brazil), Funda\c c\~ao  de Amparo \`a Pesquisa do Estado do Rio de
Janeiro (Faperj) and SR2-UERJ .

\newpage
\appendix{{\centerline{\bf{Appendix  A}}}}

\renewcommand{\theequation}{{A}.\arabic{equation}}\setcounter{equation}{0}

\noindent In order to evaluate the $C$-number contribution to the number operator, we start introducing

\be (\theta^{-})^{2} - (\theta^{+})^{2} =
-\frac{2\gamma}{|E(k)|}, \label{ap1}\ee
into (\ref{eq113}), which leads to

\bear N_{C}^{f} &=& \int_{x,k}\frac{f}{2}\left[ -\frac{2\gamma}{|E(k)|}
(u_{k}^{\ast}u_{k} - v_{k}^{\ast}v_{k}) + \frac{2\gamma}{|E^{0}|}
(u_{k}^{0\ast}u_{k}^{0} - v_{k}^{0 \ast}v_{k}^{0})\right] \nonumber
\\ &+& \frac{f}{2}\left[-\frac{2\gamma}{|E_{B}|}(u_{B}^{2}-v_{B}^{2})\right]. \label{ncmais}\ear
We are interested here in the limit $L\longrightarrow\infty$ followed by
$f\longrightarrow 1$. For the bound-state case, the function
$u_{B}$  goes to $0$ for any finite argument. The only contribution comes from $v_{B}$
resulting in $\frac{1}{2}$ in the limit of the last integral.

A much more careful treatment is required to the continuum. Let us consider here the $square$ profile functions for the $\phi(x)$ soliton anti-soliton configuration

\begin{displaymath}
\label{defsol} \phi(x)= \left\lbrace
\begin{array}{ll}
   +\phi_{0} &, x<0 \\
   -\phi_{0} &, 0\leq x \leq L \\
   +\phi_{0} &, L<x\\
  \end{array} \right.
\end{displaymath}

We can proceed as in \cite{JKKS} and construct one explicit solution $u_{k,L,\phi_{0}}(x)$, with the
requirements

\be
\int_{-\infty}^{\infty}u_{k,L,\phi_{0}}^{\ast}(x)u_{k^{'},L,\phi_{0}}(x)dx
= \frac{1}{2}\delta (k - k^{'}), \label{norm} \ee

\be \lim_{\phi_{0}\longrightarrow0} u_{k,L,\phi_{0}}(x)=
\frac{1}{2\sqrt{\pi}}e^{ikx} = u_{k}^{(0)}(x) \label{uphi0}\ee
and

\be \lim_{L_{0}\longrightarrow0} u_{k,L,\phi_{0}}(x)=
\frac{1}{2\sqrt{\pi}}e^{ikx}. \label{ul0}\ee

Indeed, for instance in the case $x<0$, we obtain:
\be
u_{k,L,\phi_{0}}(x)= = \eta_{k}^{'}e^{ikx} +
\xi_{k}^{'}e^{-ikx}, \label{defutod}\ee with \bear \eta_{k}^{'}
&=& \frac{1}{\sqrt{2}}[\eta_{k} +
i\frac{1-i\lambda}{\sqrt{1+\lambda^{2}}}e^{ikL}\xi_{-L}],
\nonumber \\
 \xi_{k}^{'} &=& \frac{1}{\sqrt{2}}[\xi_{k} +
i\frac{1-i\lambda}{\sqrt{1+\lambda^{2}}}e^{ikL}\eta_{-L}],\label{defeta}\ear
Here
\bear \eta_{k}&=&\frac{1}{2\sqrt{\pi}},\nonumber \\
\xi_{k}&=&
-\frac{i}{2\sqrt{\pi}}\frac{1-i\lambda}{[1+2\lambda^{2}+2\lambda\sqrt{1+\lambda^{2}}\cos(kl)]}[\frac{\lambda^{2}}{\sqrt{1+\lambda^{2}}}\exp(-ikL)
+ \frac{1+\lambda^{2}}{\sqrt{1+\lambda^{2}}}\exp({ikL})  + 2\lambda ].\ear
Similar expressions hold for $0\leq x\leq L $ and for $L\leq x$.

We are now able to analyse the limit $L\longrightarrow\infty$ taking into account that rapid oscilating functions involved can be seen as  distributions theories. For instance,

\bear \lim_{L\longrightarrow\infty}\frac{1}{a+b\cos(kL)}&=&
\frac{1}{\sqrt{a^{2} + b^{2}}} \nonumber \\
\lim_{L\longrightarrow\infty}\frac{\exp(\pm
ikL)}{a+b\cos(kL)}&=&\frac{1}{b} - \frac{a}{b\sqrt{a^{2} - L^{2}}}.
\label{limitsfun}\ear
After some computations, in this approach, we obtain

\be
\lim_{L\longrightarrow\infty} |u_{k,L,\phi_{0}}|^{2} \Longrightarrow
|u_{k}^{\os}|^{2},\ee
where $u_{k}^{\os}$ corresponds to the solution of the problem with
1-antisoliton profile,

\begin{displaymath}
\label{theta} \phi(x)= \left\lbrace
\begin{array}{ll}
   +\phi_{0} &, x<0 \\
   -\phi_{0} &, x>0 \\
  \end{array} \right.
\end{displaymath}
 For the $v_{k,L,\phi_{0}}(x)$, similar results can be
obtained. If we follow the same steps, we can see that

 \be
\lim_{L\longrightarrow\infty}|u_{k,L,\phi_{0}}(x)|^{2}\neq
|\lim_{L\longrightarrow\infty}u_{k,L,\phi_{0}}(x)|^{2}.
\label{inequal}\ee
Indeed
$\lim_{L\longrightarrow\infty}u_{k,L,\phi_{0}}(x)$ results in
solutions of the 1-soliton profile that are not orthonormalized. That
limit of the square is not the square of the limit, while not
entirely clear on physical grounds, is understandable since the
product of distributions is not in general well defined.

Let us face now the problem of solving (\ref{ncmais}). The
convergence factor $f(x)$ allows one to interchange the space
integrals with the limit $L\longrightarrow\infty$. We obtain that

\bear \limj N_{C}^{f}&=&\lim_{f\longrightarrow 1}
\int_{x,k}\frac{f}{2}\left[\lim_{L\longrightarrow\infty}\left(
-\frac{2\gamma}{|E_{k}|}(u_{k,L,\phi_{0}}^{\ast}u_{k,L,\phi_{0}} -
v_{k,L,\phi_{0}}^{\ast}v_{k,L,\phi_{0}})\right)\right] +{1\over 2} \nonumber \\
&=&\lim_{f\longrightarrow
1}\int_{x,k}\frac{f}{2}\left[-\frac{2\gamma}{|E_{k}|}(u_{k}^{\ast
\os}u_{k}^{\os}-v_{k}^{\ast \os}v_{k}^{\os})\right] +{1\over 2}\nonumber \\
&=&  \frac{1}{2} \int_{x,k}\left[-\frac{2\gamma}{|E_{k}|}(u_{k}^{\ast
\os}u_{k}^{\os}-v_{k}^{\ast \os}v_{k}^{\os})\right]+{1\over 2}.\ear

\newpage
\appendix{{\centerline{\bf{Appendix  B}}}}

\renewcommand{\theequation}{{B}.\arabic{equation}}\setcounter{equation}{0}

If we remember that, to the case $\gamma\neq 0$ and a one-anti-soliton background
profile, there is only the negative energy ($E_B=-\gamma$),
the bound-state field is

$$\label{solutionb2a}
\Psi_b^-(x)=\left(\begin{array}{c} 0\\ -\sqrt{2} v_B(x)
\end{array}\right).
$$
The number operator becomes

\bear  N_f&=&\sum\!\!\!\!\!\!\!\!\!\int_{x,k,k^\prime}
\left[e^{-i(E_k-E_{k^\prime})t}\left[a_k^\dagger
a_k(\theta^+\theta^{\prime +} u_{k^\prime}^{*\os}u_k^{\os}
+\theta^-\theta^{\prime -} v_{k^\prime}^{*\os}v_k^{\os})-b_k^\dagger
b_k(\theta^-\theta^{\prime -} u_{k^\prime}^{*\os}u_k^{\os}
+\theta^+\theta^{\prime +}
v_{k^\prime}^{*\os}v_k^{\os})\right]\right.\nonumber\\
&&\hskip 2cm\left.+e^{-i(E_k+E_{k^\prime})t}\left[b_k
a_k(\theta^+\theta^{\prime -} u_{k^\prime}^{\os} u_k^{\os}
-\theta^-\theta^{\prime +}
v_{k^\prime}^{\os}v_k^{\os})\right]\right.\nonumber\\&& \hskip
2cm\left.+e^{+i(E_k+E_{k^\prime})t}\left[a_k^\dagger
b_k^\dagger(\theta^-\theta^{\prime +}
{u_{k^\prime}}^{*\os}u_k^{*\os}-\theta^+\theta^{\prime
-}{v_{k^\prime}}^{*\os}v_k^{*\os})\right]\right.\nonumber\\&& \hskip
2cm\left.+\frac
12\delta(k-k^\prime)\left[({\theta^-}^2-{\theta^+}^2)u_k^{*\os}u_k^{\os}
-({\theta^-}^2-{\theta^+}^2)v_k^{*\os}v_k^{\os}\right]
\right]\nonumber\\
&=&\int_{k} (a^\dagger_ka_k-b^\dagger_kb_k)-b^\dagger_Bb_B{ +\frac
12}-\int_{k,x}
\frac{\gamma}{|E(k)|}(u^{*\os}_ku^{\os}_k-v^{*\os}_kv^{\os}_k).\label{v44} \ear
Now the c-number contribution can be computed for instance by using
the explicit form of the continuum functions $u^{\os}_k$ and
$v^{\os}_k$.

\be \label{v43}
|u_{k}^{\os}|^{2}-|v_{k}^{\os}|^{2} =
\frac{\phi_{0}k}{2\pi(\phi_{0}^{2} + k^{2})}\sin(2kx).
 \ee

 Using equations (\ref{v43}, \ref{v44}) and performing first the integration over $x$ and recognizing that the oscilating terms in
infinity do not contribute and then integrating in $k$ we find that

$$
N_f=\int_{k} (a^\dagger_ka_k-b^\dagger_kb_k)-b^\dagger_Bb_B+\frac {1}{\pi} \tan ^{-1}(\frac {\phi_0}\gamma).
$$

\end{document}